\documentclass[%
reprint,
superscriptaddress,
amsmath,amssymb,
aps,
prl,
]{revtex4-1}
\usepackage{graphicx}
\usepackage{dcolumn}
\usepackage{bm}
\usepackage{amsmath}
\usepackage{hyperref}
\hypersetup{colorlinks=true,linkcolor=blue,anchorcolor=blue,citecolor=blue,urlcolor=blue}
\usepackage{lineno}
\usepackage{enumerate}
\usepackage{color}

\begin{document}
\preprint{APS/123-QED}

\title{Laboratory-scale Perpendicular Collisionless Shock Generation and Ion Acceleration in Magnetized Head-on Colliding Plasmas}

\author{P. Liu}
\affiliation{Institute for Fusion Theory and Simulation, School of Physics, Zhejiang University, Hangzhou 310058, China}%

\author{D. Wu}
\email{dwu.phys@sjtu.edu.cn}
\affiliation{Key Laboratory for Laser Plasmas and School of Physics and Astronomy, Collaborative Innovation Center of IFSA (CICIFSA), Shanghai Jiao Tong University, Shanghai 200240, China}%

\author{D. W. Yuan}
\affiliation{Key Laboratory of Optical Astronomy, National Astronomical Observatories, Chinese Academy of Sciences, Beijing 100101, China}%

\author{G. Zhao}
\affiliation{Key Laboratory of Optical Astronomy, National Astronomical Observatories, Chinese Academy of Sciences, Beijing 100101, China}%

\author{Z. M. Sheng}
\affiliation{Institute for Fusion Theory and Simulation, School of Physics, Zhejiang University, Hangzhou 310058, China}%

\author{X. T. He}
\affiliation{Institute for Fusion Theory and Simulation, School of Physics, Zhejiang University, Hangzhou 310058, China}%

\author{J. Zhang}
\email{jzhang@iphy.ac.cn}
\affiliation{Key Laboratory for Laser Plasmas and School of Physics and Astronomy, Collaborative Innovation Center of IFSA (CICIFSA), Shanghai Jiao Tong University, Shanghai 200240, China}%

\date{\today}

\begin{abstract}
Magnetized collisionless shocks drive particle acceleration broadly in space and astrophysics. We perform the first large-scale particle-in-cell simulations with realistic laboratory parameters (density, temperature, and velocity) to investigate the magnetized shock in head-on colliding plasmas with an applied magnetic field of tens of Tesla. 
It is shown that a perpendicular collisionless shock is formed with about fourfold density jump when two pre-magnetized flows collide. This shock is also characterized by rapid increase of neutron yield, triggered by the beam-beam nuclear reactions between injected deuterons and ones reflected by the shock.
Distinct from the shocks arising from the interaction of injected flows with a magnetized background, the self-generated magnetic field in this colliding plasmas experiences a significant amplification due to the increasing diamagnetic current, approximately 30 times of upstream magnetic field. 
Moreover, we find that ions, regardless of whether they pass through or are reflected by the shock, can gain energy by the shock surfing acceleration, generating a power-law energy spectrum.
In addition, we also demonstrate that the shock mediated only by filamentation instability cannot be generated under the prevailing unmagnetized experimental parameters. 
These results provide a direct connection of astrophysical field amplification to the magnetized shock formation and nonthermal ion generation.
\end{abstract}

\maketitle
$ Introduction $.---It is widely believed that the seed fields in the Universe first generated by various fluid and plasma-kinetic instabilities, including the Kelvin-Helmholtz instability \cite{Alves_2012,Nishikawa_2014,Grismayer_PRL2013} at the shear surface of astrophysical jets and the Weibel-type instabilities \cite{WeibelPhysRevLett1959,Fried1959} in collisionless shocks, subsequently amplified by the turbulent dynamo \cite{Ryu2012,Donnert2018,Sironi_PRL2023}.
Therefore, the background magnetic fields are present in various astrophysical scenarios, such as solar wind \cite{Chen_APJ2020}, pulsar magnetospheres \cite{Comisso_PRL2018}, which usually interact with interstellar medium (ISM) \cite{Yao_MRE2021,Yao_NP2021}, forming so-called colliding plasma systems, in which magnetized collisionless shocks are generated due to pre-existing magnetic fields in the upstream plasma, which holds important astrophysical implications for the generation of high-energy cosmic particles.

As tens of Tesla magnetic fields are available in experiments, the exploration on magnetized collisionless shocks is also favored by laboratory astrophysics, in which magnetized collisionless shocks are usually created by the interaction of laser-ablated expanding plasma jets and a magnetized ambient plasma. Thanks to this setup, numerous research groups have made significant progress in experimentally exploring and numerically simulating magnetized collisionless shocks \cite{Schaeffer_PoP2012,Schaeffer_PRL2017,Schaeffer_PRL2019,Schaeffer_PoP2020,Yao_MRE2021,Yao_NP2021,Matsukiyo_PRE2022}.
Schaeffer $et \ al$. have demonstrated in detail the formation and evolution a high-Mach-number magnetized collisionless shock, along with a direct observation of particle dynamics \cite{Schaeffer_PRL2017,Schaeffer_PRL2019}. Yao $et \ al$. have studied the characteristics of a quasiperpendicular magnetized collisionless shock and associated proton acceleration \cite{Yao_MRE2021,Yao_NP2021}.

Another configuration generates collisionless shocks through laser-driven head-on colliding plasma flows in OMEGA and NIF experiments \cite{ParkPoP2015,FiuzaNP2020}. Recently, Fiuza $et \ al$. have reported a collisionless shock and electron acceleration related to the turbulence in the interpenetrating plasma experiment \cite{FiuzaNP2020}. To the best of our knowledge, this work failed to observe the jumps of ion temperature and density, as well as ion energization or reflection, which are crucial for collisionless shock characterization. Therefore, the ion kinetic behaviors in the collisionless shocks driven by such large-scale colliding plasmas are far from completely being understood, especially, the physical exploration associated with the perpendicular collisionless shock and ion acceleration in the pre-magnetized colliding plasmas remains unreported. 

\begin{figure}[t]
  \includegraphics[scale=0.38]{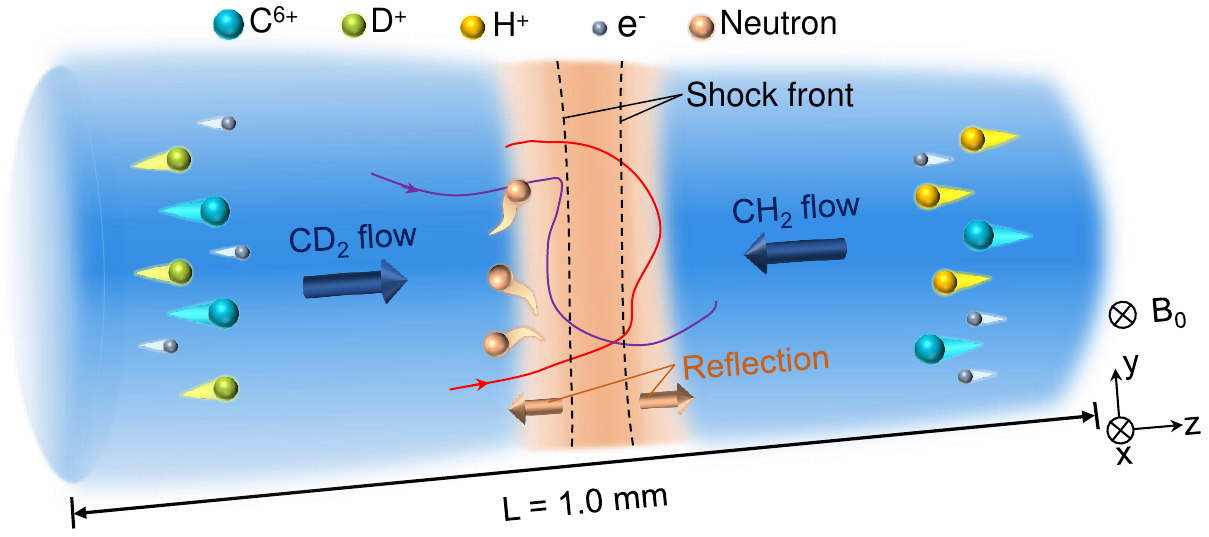}
  \caption{\label{fig:figure1} Sketch of shock structure and representative trajectories of deuterons (red and purple lines) obtained from the colliding plasma flows.}
\end{figure}

 In this Letter, we initially demonstrate through large-scale high-order implicit particle-in-cell (PIC) simulations that the shock mediated only by filamentation instability cannot be produced under the prevailing experimental parameters. However, we find that a supercritical perpendicular collisionless shock can be generated in the midplane when two pre-magnetized deuterated polyethylene ($ \mathrm{CD_2} $) and polyethylene ($ \mathrm{CH_2} $) flows collide, as illustrated in Fig. \ref{fig:figure1}, in which the both the reflected (purple line) and the transmitted(red line) particles are accelerated by the convective electric field. This shock is self-consistently manifested in neutron diagnosis, showing that the deuterium-deuterium (D-D) beam-beam reactions triggered by this magnetized shock lead to a substantial enhancement of neutron yield.

To get a deeper insight of the collisionless shock generation and associated ion kinetic behaviors in both unmagnetized and magnetized colliding plasmas, we perform a series of large-scale two-dimensional (2D) PIC simulations by employing the high-order implicit LAPINS code \cite{Wu2019,Wu2021}, which includes the nuclear reaction module \cite{Wu2021}, serving as an important diagnostic tool for current experiments \cite{RossPhysRevLett2017,HigginsonPOP2019}. The realistic beam velocity and ion-to-electron mass ratio are used in our simulations, which are different from the transformation PIC simulations in Refs. \cite{ParkPoP2015,FiuzaNP2020}. Although this method can subtly reduce computational burden, the impacts of nonphysical mass ratio on ion filamentation dynamics cannot be overlooked at the nonlinear stage \cite{RuyerPhysRevLett2016}, and collisional thermalization and nuclear reaction processes cannot be realized self-consistently.

$ Simulation \ setups $.---In our simulations, the $ \mathrm{CD_2} $ and $ \mathrm{CH_2} $ flows are injected from the left and right sides of the simulation box, respectively. The beam parameters closely resemble those employed in NIF and OMEGA experimental setups \cite{KuglandNP2012,FoxPhysRevLett2013,HuntingtonNP2015,ParkPoP2015,HuntingtonPoP2017,RossPhysRevLett2017,FiuzaNP2020,HigginsonPOP2019}, featuring an velocity of $v_0 = 1000\ \text{km/s}$, a density of $n_e = 4 \times 10^{19}\ \text{cm}^{-3}$, and a temperature of $T_i = T_e = 100 \ \mathrm{eV}$.
The plasma flows are weakly magnetized with an ambient magnetic field $\mathbf{B}_0 =B_0\mathbf{\hat{e}}_x$ ($\vartheta_{Bn} = 90 ^{\circ} $), where $B_0=20 \ \mathrm{T}$, which is available for current laboratories \cite{Schaeffer_PoP2012,Schaeffer_PRL2017,Schaeffer_PRL2019,Schaeffer_PoP2020,Yao_MRE2021,Yao_NP2021,Matsukiyo_PRE2022,Fujioka2013,Law2016}.
The simulation window is $ 1.0 \ \mathrm{mm} \ (z) \times 0.4 \ \mathrm{mm} \ (y) $ with resolution of $ dz=dy= 2.0\ \mu$m ($\simeq 0.02 c/\omega_{pi} $), where $c$ is light speed in vacuum and $\omega_{pi}$ is the plasma frequency of ions. Periodic and absorbing boundary conditions are adopted in the $y$ and $z$ directions, respectively.

\begin{figure}[t]
  \includegraphics[scale=0.58]{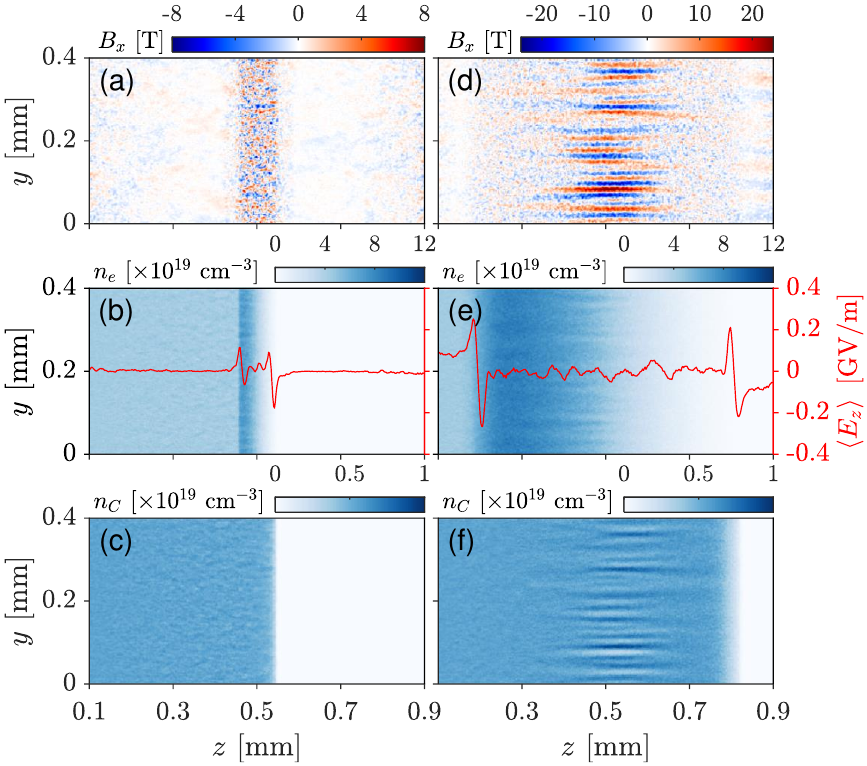}
  \caption{\label{fig:figure2} Left column: $t = 520$ ps, right column: $t = 800$ ps. Snapshots of the self-generated magnetic field $B_x$ [(a),(d)], electron density $n_e$ [(b),(e)], and carbon density $n_C$ [(c),(f)]. The red lines in (b) and (e) are the longitudinal electric field $\left< E_z \right> $ profiles at the corresponding times, respectively. Here, $\left< \cdot \right>$ represents averaging over the $y$ direction.}
\end{figure}

$ Unmagnetized \ scenario $.---When two unmagnetized plasma flows collide, the ion filamentation instability (IFI) is easily triggered, generating a magnetic field $ B_x $ with several Tesla at the early stage of linear phase [see Fig. \ref{fig:figure2}(a)]. This self-generated magnetic field is capable of trapping the injected electrons due to their gyro-radius of $r_{ce} \simeq 9.5 \times 10^{-4} \ \mathrm{mm}$, yet it fails to capture ions, which therefore leads to a longitudinal electrostatic field [see red line in Fig. \ref{fig:figure2}(b)] characterized by charge separation. As a result, a portion of trapped electrons can be accelerated by this electrostatic field and move along the opposite ($-z$) direction, while the ion beams interpenetrate into each other, as shown in Fig. \ref{fig:figure2}(c). The situations for $\mathrm{CH_2}$ flow are quite similar to those for $\mathrm{CD_2}$ and therefore are not shown.
We see that, from $ t= 520$ ps to $800$ ps, the electrostatic field and accelerated electrons move synchronously along the $-z$ direction, leading to a doubling of electron density, and $B_x$ in the flow front is very small, which is distinct from the collisionless shocks mediated by the IFI \cite{Kato_2008,BohdanPRL_FieldAmp}. The similar results are reported by Stockem $et \ al$. \cite{Stockem_PRL2014}, however, they failed to observe the reverse motion of electrons in their simulations.

As the ion beams further interpenetrate, at the moment of $t=800$ ps, the self-organized flow-aligned filamentous magnetic field starts to merge and tends to saturate, and its peak strength is $B_x \simeq 20\ \mathrm{T}$ [see Fig. \ref{fig:figure2}(d)], which can be well explained by the linear dispersion theory of ion filamentation mode, given by
\begin{equation}\label{Eq:1}
\begin{aligned}
   k^{2}c^2=\omega ^2 & -\omega _{pe}^{2}\left[ 1-\mathcal{W}\left( \zeta _e \right) \right] \\
   &- \omega _{pi}^{2}\left[ 1-\left( A_i+1 \right) \mathcal{W}\left( \zeta _i \right) \right],
\end{aligned}
\end{equation}
which is written in the form of $\mathcal{W}$ function, $ \mathcal{W}\left( \zeta_\alpha \right) =\pi ^{-1/2}\ \int_{-\infty}^{\infty}{x\exp \left( -x^2 \right) /\left( x-\zeta_\alpha \right)}\mathrm{d}x $ with $\zeta_\alpha = \omega /k v_{t\alpha,\perp}$ for the $\alpha$th charged particle, $A_{i}=\left( 2v_{0}^{2}+v_{ti ,\parallel}^{2} \right)/v_{ti ,\perp}^{2}-1$,  where $v_{ti ,\parallel}$ and $v_{ti ,\perp}$ are respectively the parallel and perpendicular thermal velocities of ions with respect to flow velocity $v_{0}$. The growth rate of IFI predicted by Eq. (\ref{Eq:1}) is $ \Gamma_{\mathrm{max}} \simeq 1.8 \times 10^{-3}\omega _{pi} $, and corresponding saturated field is $ B_{\mathrm{sat}} \simeq \Gamma_{\mathrm{max}}^2 m_i c / q_i v_0 k_{\mathrm{sat}}\simeq 14 \ \mathrm{T}$ with $k_{\mathrm{sat}} \sim 2\pi \omega_{pi}/c$ \cite{Davidson1972}. These results can provide possible explanations that the OMEGA experiment observed a filamentous magnetic field within a nanosecond time scale \cite{HuntingtonNP2015}, and the NIF experiment failed to observed the ion density compression related to hydrodynamic jump condition \cite{FiuzaNP2020}. 

\begin{figure}[b]
  \includegraphics[scale=0.57]{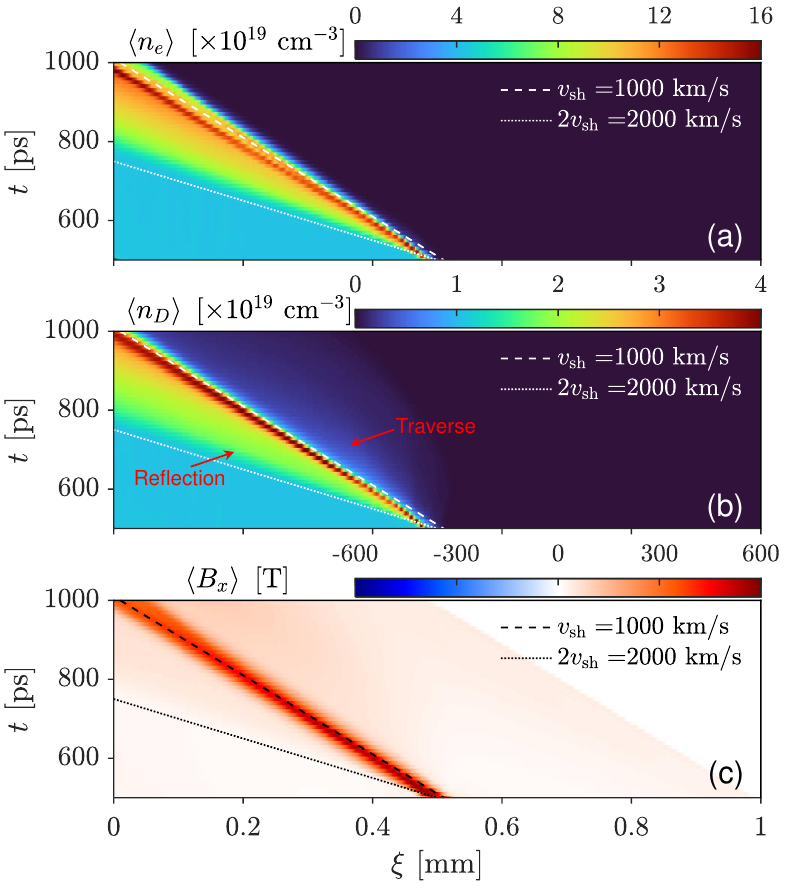}
  \caption{\label{fig:figure3} Pre-magnetized colliding plasmas. Spatiotemporal structures of electron density (a), deuteron density (b), and self-generated magnetic field (c). The dashed lines depicts the trajectory of the shock front with the velocity of $v_{\mathrm{sh}}$, and the dotted lines represents the trajectory of reflected particles with the velocity of $2v_{\mathrm{sh}}$; $\left< \cdot \right>$ represents averaging over the $y$ direction. Here, $\xi=z-v_0(t-t^ \prime)$ when $t \geq t^ \prime$ with $t^ \prime = 500$ ps being the moment of colliding between two flows.}
\end{figure}

\begin{figure}[t]
  \includegraphics[scale=0.61]{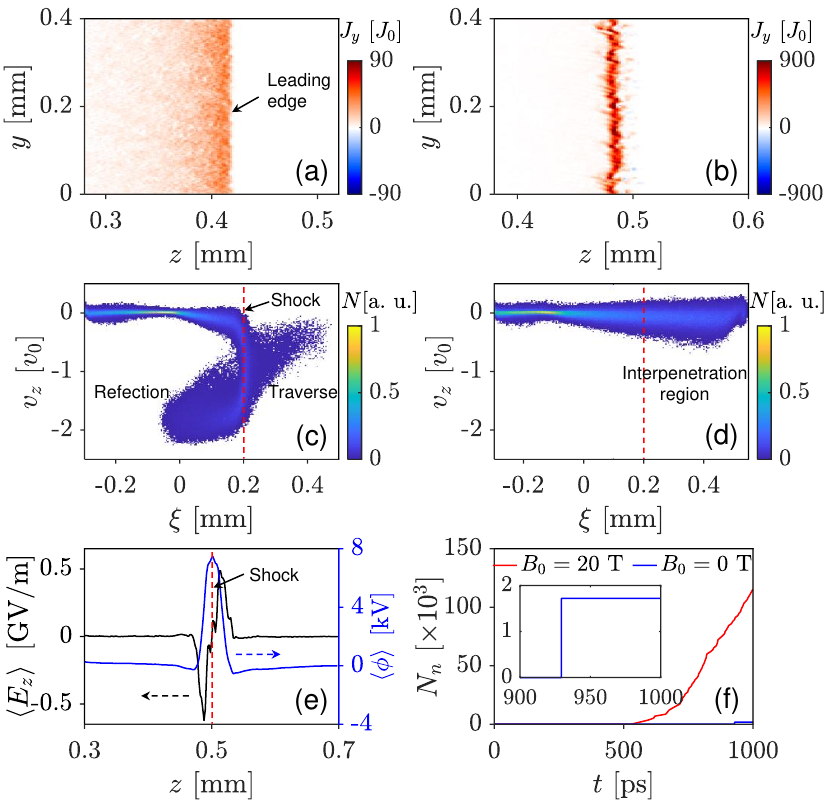}
  \caption{\label{fig:figure4} [(a),(b)] Snapshots of the diamagnetic current at two different moments: $t = 400$ ps (a) and $t = 500$ ps (b) in the magnetized scenario, where $J_0 = e c n_c $ with $n_c = 1.1 \times 10^{15} \ \mathrm{cm}^{-3}$.
  [(c),(d)] Phase space $\xi - v_z$ distributions of deuterons in the magnetized (c) and unmagnetized (d) cases at $t = 800$ ps, where $\xi=z-v_0(t-t^ \prime)$ with $t^ \prime = 500$ ps. 
  (e) Longitudinal electric field $\left < E_z \right>$ and electric potential $\left < \phi \right>$ profiles at $t = 520$ ps in the magnetized scenario, where $\left< \cdot \right>$ represents averaging over the $y$ direction.
  (f) Neutron yields as the functions of time in the magnetized (red) and unmagnetized (blue) cases, where the inset is the zoomed-in image of the corresponding part.}
\end{figure}

$ Magnetized \ shock \ generation$.---When two pre-magnetized plasma flows collide, a supercritical perpendicular shock wave is produced with a velocity $v_{\mathrm{sh}} \simeq 1000.0 \ \mathrm{km/s}$ in the reference frame of upstream flow. This can be characterized by the temporal evolution of the peak particle densities and magnetic field, as illustrated by dashed lines in Fig. \ref{fig:figure3}, and corresponding magnetosonic Mach number is 
\begin{equation}\label{Eq:2}
   M_{\mathrm{ms}}= \frac{v_{\mathrm{sh}}}{\sqrt{v_A^2+c_s^2}} \simeq 9,
\end{equation}
where $v_A=B_0/\sqrt{\mu_0(m_e n_e + m_i n_i)}\simeq 48.8 \ \mathrm{km/s}$ is upstream Alfv\'{e}nic speed, $c_s = \sqrt{(T_D+T_e)/m_D} \simeq 97.8 \ \mathrm{km/s}$ is sound speed of upstream deuterons. Figs. \ref{fig:figure3}(a) and \ref{fig:figure3}(b) show typical structures of a magnetized collisionless shock with a well-defined density compression (jump) for both electrons and deuterons, $\delta = n_{\mathrm{ds}}/n_{\mathrm{us}} \simeq 3.8$, which is close to the upper limit of the prediction of the Rankine-Hugoniot equations, where $n_{\mathrm{us}}$ and $n_{\mathrm{ds}}$ are the upstream and downstream densities of the shock.
It is of interest that the downstream magnetic field can reach several hundreds of Tesla in the overshoot zone, approximately 30 times of upstream magnetic field, which shows a significant difference from the configurations of  expanding piston plasmas interacting with a magnetized background \cite{Schaeffer_PoP2012,Schaeffer_PRL2017,Schaeffer_PRL2019,Schaeffer_PoP2020,Yao_NP2021,Yao_MRE2021}, where magnetic field is only a few times amplified. This is explained as follows, in the presence of a applied magnetic field, the plasma flows can sweep out the field due to the frozen-in effect \cite{Schaeffer_PoP2020}, compressing the magnetic field near the leading edge of the flow and therefore creating an increasing diamagnetic current $\mathbf{J}_y \propto \left( \mathbf{B}_x \times \nabla n \right)$ in Figs. \ref{fig:figure4}(a) and \ref{fig:figure4}(b), where $\nabla n$ is the density gradient of the flow leading edge, and the flow from the other side contributes a current with the opposite direction (not shown). This diamagnetic current, in turn, creates a magnetic compression [see Fig. \ref{fig:figure3}(c)], which is the underlying factor responsible for the shock generation, and it is strongly associated with the field amplification in astrophysics \cite{BohdanPRL_FieldAmp,Bell_TrubAmp,PetersonPRL2021,Peterson_2022}. 

Note that the ion Larmor radius in $\sim 500$ T magnetic fields is $r_{ci} \sim 42 \ \mu m$, which is larger than the shock transition $\Delta_{\mathrm{sh}} \sim 25\  \mu m$, therefore, deuterons could have traversed shock and penetrated into downstream regions. However, only a portion of ions pass through the shock, which is due to the emergence of the longitudinal bipolar electrostatic field $E_z$ and corresponding cross-shock electrostatic potential $\phi$ in the shock front region [see Fig. \ref{fig:figure4}(e)], most of the incoming ions are reflected back due to the velocity $v_z$ satisfying
\begin{equation}\label{Eq:3}
 \frac{1}{2} m_i v_z^2 < q_i\phi.
\end{equation}
The reflection of deuterons is also clearly shown in $\xi - v_z$ phase space, as illustrated in Fig. \ref{fig:figure4}(c). 
Note that in a marked distinction from the magnetized scenario, the longitudinal electrostatic field in the unmagnetized scenario exhibits a precisely opposed direction, as expected, the ions are not reflected longitudinally in this case [see Fig. \ref{fig:figure4}(d)].

To further confirm the generation of the shock, we also analyze the temporal evolution of neutron yield from D-D reactions $ \mathrm{D}(\mathrm{d},\mathrm{n}) ^3\mathrm{He}\ \left( Q=3.269\ \mathrm{MeV} \right)$ \cite{HigginsonPOP2019,Atzeni_2004}. At the first stage, i.e., prior to the colliding of the two plasma flows, the neutrons can only originate from thermonuclear reactions in the single injected $ \mathrm{CD}_2 $ flow, and it is found that neither magnetized nor unmagnetized scenarios result in neutron production, as shown in Fig. \ref{fig:figure4}(f). After the colliding of two flows, we see that only a minute quantity of neutrons is generated in the unmagnetized case. However, the neutrons increase rapidly in the magnetized scenario, which is because the D-D beam-beam reactions from injected and reflected flows are dominant and its average reactivity is significantly higher than that of thermonuclear.
Therefore, the sudden increase of neutron yield can be used as a direct evidence for the generation of magnetized collisionless shock.

\begin{figure}[t] 
  \includegraphics[scale=0.59]{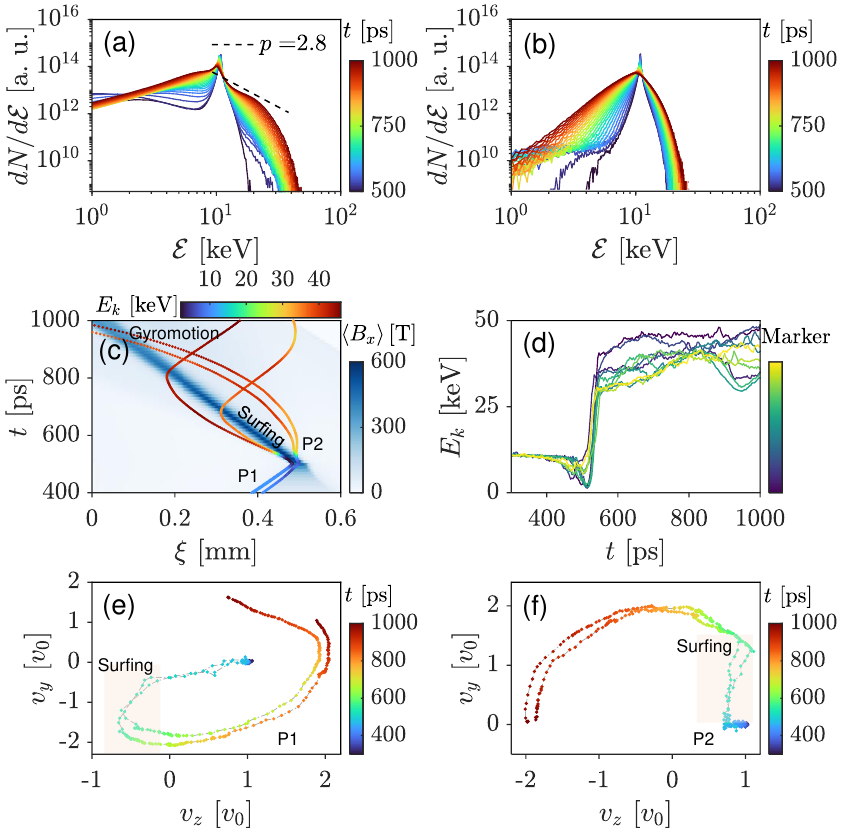}
  \caption{\label{fig:figure5} [(a),(b)] Temporal evolution of the deuteron energy spectra (colored by time) for the magnetized (a) and unmagnetized (b) scenarios. Note that the shock accelerates deuterons to $\sim 50 \ \mathrm{keV} $, leading to the formation of a non-thermal power-law tail $\mathcal{E}^{-p}$ with spectral index $p \simeq 2.8$ (black dashed line). 
  (c) Trajectories of four deuterons which are divided into two populations (P1 and P2 are represent the reflected and transmitted ones, respectively) in the $\xi-t$ diagram, the colour maps on the top and right represent the deuteron energy and magnetic field, respectively.
  (d) Evolution of the kinetic energies (colored by particles) for 10 representative particles selected form the non-thermal tail of (a).
  [(e),(f)] The $v_z - v_y$ spaces (colored by time) of P1 (e) and P2 (f), the grey shaded areas correspond to the surfing stage.}
\end{figure}

$ Ion \ acceleration$.---In contrast to the unmagnetized case, the deuterons in the magnetized scenario gain energy through the convective electric fields $\mathbf{E}_c = -\mathbf{v} \times \mathbf{B} $ in the shock stationary frame when trapped around the shock, and their energy spectrum evolution is presented in Fig. \ref{fig:figure5}(a). A pronounced nonthermal tail develops, displaying a power-law energy $ dN/d\mathcal{E} \propto \mathcal{E}^{-p}$ with spectral index $p \simeq 2.8$. It is remarkably distinct from unmagnetized counterpart in Fig. \ref{fig:figure5}(b), which shows a quasi-thermal distribution for the high-energy deuterons due the stochastic energization and collisional thermalization, although the ions in the interpenetrating region [$\xi \geq 0.2 \ \mathrm{mm}$ in Fig. \ref{fig:figure4}(d)] gain a minor amount of energy from the electromagnetic field. The high-energy cutoff is also enhanced by a factor exceeding 2 due to the existence of the magnetized shock.

To unveil the particle acceleration mechanisms, we trace the trajectories of a sample of $4 \times 10^6$ deuterons, it is found that $\sim$ 0.5\% of them ultimately populate the non-thermal tail of energy spectrum. In Fig. \ref{fig:figure5}(c), we show the energy evolution and trajectories of four representative deuterons, which are divided into two populations: reflected (P1) and transmitted (P2) particles. 
For the former (P1), they are first decelerated and reflected by the electrostatic potential at the shock front ($t \simeq 500\ \mathrm{ps}$) due to satisfying Eq. (\ref{Eq:3}), and they are then accelerated via the convective electric field $\mathbf{E}_c = -v_z B_x \mathbf{\hat{e}}_y $ of incoming flow while confined around the shock front. This way of gaining energy is a typical shock surfing acceleration (SSA) \cite{SSA_Hoshino,SSA_GRL,SSA_PSS}. Subsequently, under the influence of the upstream Lorentz force, the accelerated ions return to the front of the shock once again after several hundred picoseconds. At this time, the kinetic energy of ions becomes greater than the potential energy, thus they overcome the electrostatic potential barrier and penetrate toward the downstream region.

While for the latter (P2), they arrived at the colliding plane prior to the complete formation of the magnetized shock, during which time the electrostatic field remains relatively weak, insufficient to reflect the upstream ions, and therefore, these ions can directly pass through the shock and enter downstream region. We find that these transmitted ones also experience a SSA, which is attributed to the downstream convective electric field $ \mathbf{E}_c = v_z B_x \mathbf{\hat{e}}_y $ along the positive $y$ direction. It is worth noting that there is only a minor increase in $v_z$ [see Fig. \ref{fig:figure5}(f)], indicating the ion acceleration by the longitudinal bipolar electric field $E_z$ can be neglected.

No matter P1 or P2, their energy gains primarily come from transverse acceleration, as shown in Figs. \ref{fig:figure5}(e) and \ref{fig:figure5}(f). This acceleration process is a characteristic exclusive to this colliding plasma system. Instead, in the case of interaction between expanding piston plasmas and a magnetized background plasma, once ions penetrate downstream, they are incapable of maintaining further acceleration \cite{Yao_NP2021,Yao_MRE2021}.

Additionally, we show the kinetic energy evolution of 10 particles that eventually populate the non-thermal tail in Fig. \ref{fig:figure5}(d). A common feature of these tracks is
the rapid energy increase from $ \mathcal{E} \sim 10 \ \mathrm{keV} $ up to $ \mathcal{E} \sim 40 - 50 \ \mathrm{keV}$ within $\sim 50$ ps. The maximum energy of the accelerated ions can be obtained by the balance between the upstream Lorentz force and the electric field force
\begin{equation}\label{Eq:4}
 \mathcal{E}_{\mathrm{max}} \simeq \frac{1}{2} m_i \left( \frac{E_z}{B_x} \right)^2,
\end{equation}
which is in accordance with the SSA mechanism. Indeed, we have verified that the overwhelming majority of ions belonging to the high-energy tail undergo such a sudden episode of energy gain. 
These high-energy ions from SSA may be a candidate for slow pick-up ions to start diffusive shock acceleration.

We also explore how the behaviors of the magnetized shock, propelled by colliding plasmas, are influenced by the background plasma, whose density is $1$-$2$ orders of magnitude lower that of the injected flows. It is found that the physical processes related to shock after flow colliding are very similar to those in Fig. \ref{fig:figure3}, and a shock created through the interaction of injected flows and a background plasma is also observed. These results are presented in Supplemental Material \cite{Supplement}.

$ Discussion \ and \ conclusion$.---In this work, the physical evolutions pertaining to laser-ablated foil targets are not taken into account. However, during this process, the magnetic field can be generated by Biermann battery effect and frozen in the flow \cite{HuntingtonNP2015,PRL_Matteucci_Biermann}, consequently, upon colliding, the ion beams may become magnetized. This magnetized colliding flows are prone to amplifying the magnetic field originating from IFI to a turbulent field \cite{Pliu_PRL2024}, potentially leading to the formation of a collisionless shock in recent colliding plasma experiment \cite{FiuzaNP2020}.

In conclusion, we first demonstrate the collisionless shock mediated only by filamentation instability cannot be generated in the unmagnetized scenario. 
However, in the magnetized colliding plasmas with an applied magnetic field of tens of Tesla,
a perpendicular collisionless shock is produced when two pre-magnetized flows collide, and a distinct ion reflection is shown in the phase space. It is self-consistently manifested in neutron diagnosis, showing that the neutron yield is substantially enhanced due to the beam-beam reactions between the incoming deuterons and ones reflected by this magnetized shock. In this scenario, ions, regardless of whether they pass through or are reflected by the shock, can gain energy in the convective electric field, which is consistent with SSA mechanism. Our findings theoretically provide a direct connection of astrophysical field amplification to the magnetized shock formation and nonthermal ion generation \cite{BohdanPRL_FieldAmp,Bell_TrubAmp,Bohdan_2019}, which can be validated through current astrophysical experiments.

This work was supported by the Strategic Priority Research Program of Chinese Academy of Sciences (Grants No. XDA25050500 and No. XDA25010100), the National Natural Science Foundation of China (Grants No. 12075204, No. 11875235, and No. 61627901), the Shanghai Municipal Science and Technology Key Project (No. 22JC1401500), and the National Supercomputing Tianjin Center Fusion Support Program. D. W. thanks the sponsorship from Yangyang Development Fund.


\bibliography{Reference0505}

\end{document}